\documentclass[draft,JGR]{agujournal2019}
\usepackage{url} 
\usepackage{natbib}
\usepackage{lineno}
\usepackage[inline]{trackchanges} 
\usepackage{soul}

\draftfalse

\journalname{Earth and Space Science}

\usepackage{booktabs}
\usepackage{amsmath}
\usepackage{amssymb}
\usepackage{siunitx}

\newcommand{\Te}{T_\mathrm{e}}
\newcommand{\Ti}{T_\mathrm{i}}
\newcommand{\Tperp}{T_\perp}
\newcommand{\Tpar}{T_\parallel}
\newcommand{\Rsun}{R_\odot}

\newcommand{\kSps}{\mathrm{kSamples\,s^{-1}}}

\begin{document}

\title{EMBER: Machine-Learning Detection of Modulated Ion Acoustic Waves and Associated Core-Electron Heating in the Solar Wind with Parker Solar Probe}

\authors{Argyro Sasli\affil{1}, Karish Seebaluck\affil{1}, Chris Colpitts\affil{1}, and Michael Coughlin\affil{1}}

\affiliation{1}{School of Physics and Astronomy, University of Minnesota, Minneapolis, MN 55455, USA}

\correspondingauthor{Argyro Sasli}{asasli@umn.edu}

\begin{keypoints}
\item EMBER is an open source machine learning pipeline that detects modulated ion acoustic wave events in Parker Solar Probe FIELDS burst data.
\item An ensemble of 16 background-only anomaly detectors recovers approximately 93\% of labeled wave events at an estimated 1\% false alarm rate, noting that this value is approximate due to limited background statistics.
\item Intervals flagged by EMBER show core electron temperatures above adiabatic cooling and elevated electron to ion temperature ratios.
\end{keypoints}

\begin{abstract}
Modulated ion acoustic waves (IAWs) --- including triggered ion acoustic waves (TIAWs) and frequency-dispersed ion acoustic waves (FDIAWs) --- are increasingly recognized as efficient drivers of electron heating in the solar wind through nonlinear wave--particle interactions. Identification of these events in the Parker Solar Probe (PSP) FIELDS burst-mode archive has so far relied on expert visual inspection and does not scale to the full mission. We present \textsc{Ember} (Electron heating from Modulated Burst-mode Event Recognition), an open-source pipeline that converts PSP FIELDS Digital Burst Memory (DBM) voltage bursts into log-scaled Fourier spectrograms and applies a multi-detector, background-only anomaly detection suite. The suite combines physics-motivated detectors, classical outlier detectors, and deep learning detectors. The \textsc{Ember} ensemble recovers $93\%$ of the anomalous events at 1\% FAR (1 false positive per 100 held-out backgrounds). Coincident SWEAP/SPAN diagnostics show that flagged intervals exhibit core perpendicular electron temperatures above the adiabatic cooling expectation and elevated $\Te/\Ti$, reproducing the preferential-heating phenomenology established by prior manual studies without any use of electron temperatures in the detection step.
\end{abstract}

\section{Introduction}
\label{sec:intro}

Plasma is the most common state of matter in the universe, and wave--particle interactions are fundamental drivers of plasma evolution in both planetary magnetospheres and the heliosphere: in Earth's radiation belts they control the acceleration, transport, and loss of energetic particles \citep{ripoll2020,colpitts}, while in the solar wind they regulate particle heating and thermodynamic evolution \citep{Li2019,yuan,chorus}. Plasma waves may be broadly categorized into modes dominated by electric fields (electrostatic, E) and modes involving coupled electric and magnetic fields (electromagnetic, E and B). Ion acoustic waves (IAWs) are longitudinal electrostatic waves that commonly arise from plasma instabilities driven by currents, beams, and density gradients \citep{filbert,bernstein}. More broadly, modulation of higher-frequency, electron-scale waves by lower-frequency, ion- or MHD-scale waves is a widespread mechanism for coupling disparate plasma regimes, well documented in Earth's magnetosphere where ULF pulsations and electromagnetic ion cyclotron waves modulate whistler-mode chorus and related electron-scale emissions that govern radiation-belt electron dynamics \citep{bortnik2007, li2011, watt2011, colpitts2}. Of particular interest are modulated IAWs, including triggered ion acoustic waves (TIAWs), in which high-frequency ion acoustic waves are modulated by lower-frequency waves \citep{mozercore}. These modulated waves are increasingly recognized as efficient drivers of electron heating through nonlinear wave--particle interactions \citep{spanswick, Mozer_2021}.

The plasma in the Sun's corona is heated to temperatures of order \SI{e6}{\kelvin} \citep{pawsey}, and expands into interplanetary space to form the solar wind \citep{parker}. The slow solar wind ($\sim\SI{400}{\kilo\metre\per\second}$) is thought to originate predominantly from equatorial streamer-belt regions, while the fast wind ($\sim\SI{750}{\kilo\metre\per\second}$) arises from high-latitude coronal holes \citep{feldman,cranmer}. Electrons in the slow solar wind consist of core, halo, and strahl populations \citep{pierrard}, and test-particle simulations and spacecraft observations have shown that both whistler-mode waves and IAWs can produce significant electron scattering and heating in the solar wind
\citep{cattell,vo,kellog}, with modulated waves further amplifying these interactions \citep{spanswick}. Recent Parker Solar Probe (PSP) observations demonstrate that modulated IAWs are strongly correlated with enhancements in the electron core temperature and in the temperature ratio $\Te/\Ti$, indicating preferential electron heating via nonlinear wave--particle interactions \citep{mozercore,mozer2}.

PSP is the first spacecraft to directly sample the solar corona and the nascent solar wind, owing to its unprecedented proximity to the Sun, reaching distances within a few solar radii where the solar wind is formed  \citep{kasper,fields}. It provides in-situ measurements of electron and ion velocity distribution functions (VDFs) and high-cadence electric and magnetic field fluctuations, enabling direct detection of frequency-dispersed ion acoustic waves (FDIAWs)—narrowband electrostatic emissions whose central frequency chirps up or down by factors of 3--10 within a fraction of a second \citep{malaspina}—and modulated TIAWs, along with quantification of associated electron heating. However, the identification of modulated IAW events in the FIELDS Digital Burst Memory (DBM) archive has so far been performed by expert visual inspection of spectrograms and waveforms. Each 6-hour DBM block contains dozens of burst windows of $\sim\num{524288}$ samples each, and events of interest occupy only a small fraction of the archive. A scalable, reproducible detection method is a prerequisite for statistical studies of wave-driven heating across the full mission and for any downstream effort that seeks to associate these localized kinetic signatures with global coronal context (for example, via foundation-model latent embeddings such as those now being released by NASA's \textsc{Surya} Helio Foundation Model).

In this paper we introduce \textsc{Ember} (Electron heating from Modulated Burst-mode Event Recognition), an open-source Python pipeline that converts PSP FIELDS DBM voltage data into log-scaled Fourier spectrograms and applies a multi-detector, background-only anomaly detection suite. We describe the pipeline in Section~\ref{sec:data}--\ref{sec:method}, benchmark its performance on a curated catalog of 42 anomalous and 496 background DBM spectrograms from PSP Encounters~6-9 and~15 in Section~\ref{sec:results}, show in Section~\ref{sec:heating} that flagged intervals reproduce the electron-heating phenomenology established by prior manual studies, and discuss extensions to foundation-model features and coronal-context studies in Section~\ref{sec:discussion}. Section~\ref{sec:conclusions} concludes.

\section{Data}
\label{sec:data}

\subsection{Parker Solar Probe observations}

PSP \citep{fields,kasper} provides high-cadence plasma and field measurements uniquely suited to studying wave--particle interactions and electron heating in the near-Sun solar wind. PSP’s encounters sample a wide range of heliocentric distances. This study focuses on Encounters 6–9 (spanning $\sim 21\text{--}40,\Rsun$) and Encounter 15 ($\sim 12\text{--}35,\Rsun$). These progressively smaller perihelion distances are achieved through a series of gravity assists from Venus, which step the orbit inward; two such assists occurred between Encounters 6 and 15.

Electric and magnetic fields are measured by the FIELDS instrument suite \citep{fields}. High-rate burst-mode voltage data (up to $\SI{150}{\kSps}$) and onboard spectrograms from the Digital Fields Board \citep{dfb} allow identification of narrow band electrostatic waves and modulated wave packets, including FDIAWs and TIAWs. Three-dimensional electron VDFs are obtained from the SWEAP/SPAN electron analyzers \citep{kasper}, providing full angular coverage over $2\,\mathrm{eV}$--$2\,\mathrm{keV}$ at \SI{0.435}{\second} cadence; proton bulk velocity from SPAN-Ai is used to determine solar-wind speed and account for spacecraft motion.

\subsection{FIELDS DBM burst-mode products}

\textsc{Ember} ingests Level-2 FIELDS DBM CDF files from NASA CDAWeb and Berkeley SSL. PSP has four antennas mounted at the corners of its heat shield, arranged diagonally; antennas 1--2 and 3--4 form opposing pairs. Two product types are used: \emph{DVAC}, the differential voltage between opposing antennas (pair 1--2 or 3--4), and \emph{VAC}, single-ended voltages from each antenna, which are converted to differential voltage via $\mathrm{dvac}_{12} = \mathrm{vac}_1 - \mathrm{vac}_2$ and $\mathrm{dvac}_{34} = \mathrm{vac}_3 - \mathrm{vac}_4$ before spectrogram computation. Each CDF file covers a 6-hour block and contains multiple burst windows of typically $\num{524288}$ samples. Timestamps are converted from TT2000 to UTC during preprocessing.

\subsection{Hand-labeled catalog}
\label{sec:catalog}

To train and benchmark \textsc{Ember} we use a manually curated catalog drawn from PSP Encounters~6 through~9 (25~September 2020--13~August 2021) and Encounter~15 (March~2023). The catalog comprises $538$ DBM spectrograms, of which $496$ are background intervals (quiescent solar-wind turbulence or unrelated narrowband activity) and $42$ are \emph{anomalies} in two physical classes: \textbf{Label~1}, comprising modulated/coupled wave events (TIAWs and TIAW-like modulated FDIAWs) associated with strong core-electron heating, and \textbf{Label~2}, comprising FDIAWs with weak associated heating. For each anomalous event we record start time, end time, carrier and modulation frequencies, amplitude, duration, and physical class. This split (background-dominant, rare-event) is the defining constraint on the method described in Section~\ref{sec:method} and motivates the use of anomaly detection rather than supervised classification. 

\begin{figure}[ht]
    \centering
    \includegraphics[width=\linewidth]{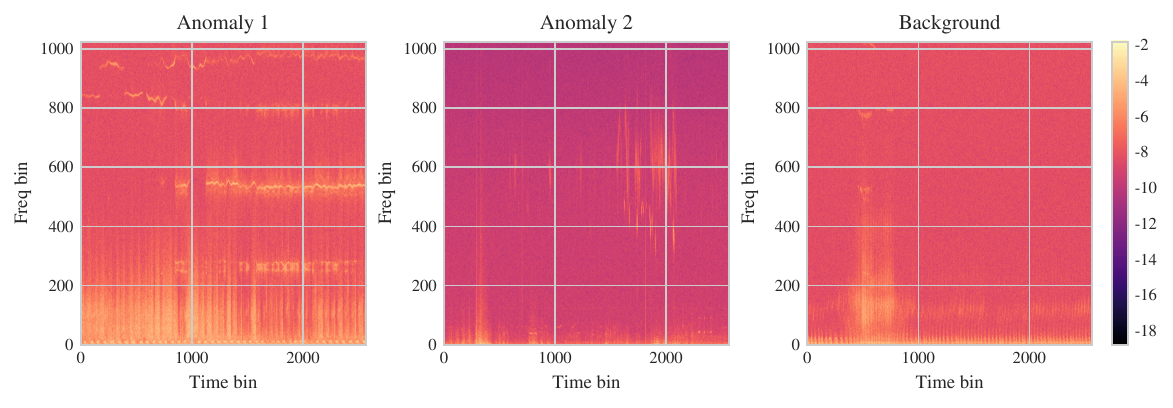}
    \caption{
    Representative PSP log-amplitude spectrograms for the three classes used in this study.
    From left to right: \textbf{Anomaly 1} (TIAW-like coupled-wave signatures), \textbf{Anomaly 2} (FDIAW--like structures), and \textbf{Background} (Label 0).
    Each panel is shown in the same preprocessing space produced by the \textsc{Ember} pipeline (\S\ref{sec:method}); brighter colors indicate higher spectral power.
    }
    \label{fig:class_examples}
\end{figure}

Figure~\ref{fig:class_examples} shows representative spectrograms for the three PSP classes after preprocessing. Anomaly 1 (TIAW-like) shows a persistent narrowband carrier near a few hundred Hz that is amplitude-modulated by a low-frequency ($\sim$1–5 Hz) envelope, producing the characteristic ladder of sidebands described in Section \ref{sec:pre}. Anomaly 2 (FDIAW-like) instead shows a single narrowband emission whose central frequency chirps rapidly up or down by factors of several over a fraction of a second, with no coherent low-frequency modulating envelope. Background intervals (Label 0) lack both signatures and are dominated by broadband solar-wind turbulence and instrument noise.

\section{The \textsc{Ember} pipeline}
\label{sec:method}

\begin{figure}[t]
\centering
\includegraphics[scale=0.4]{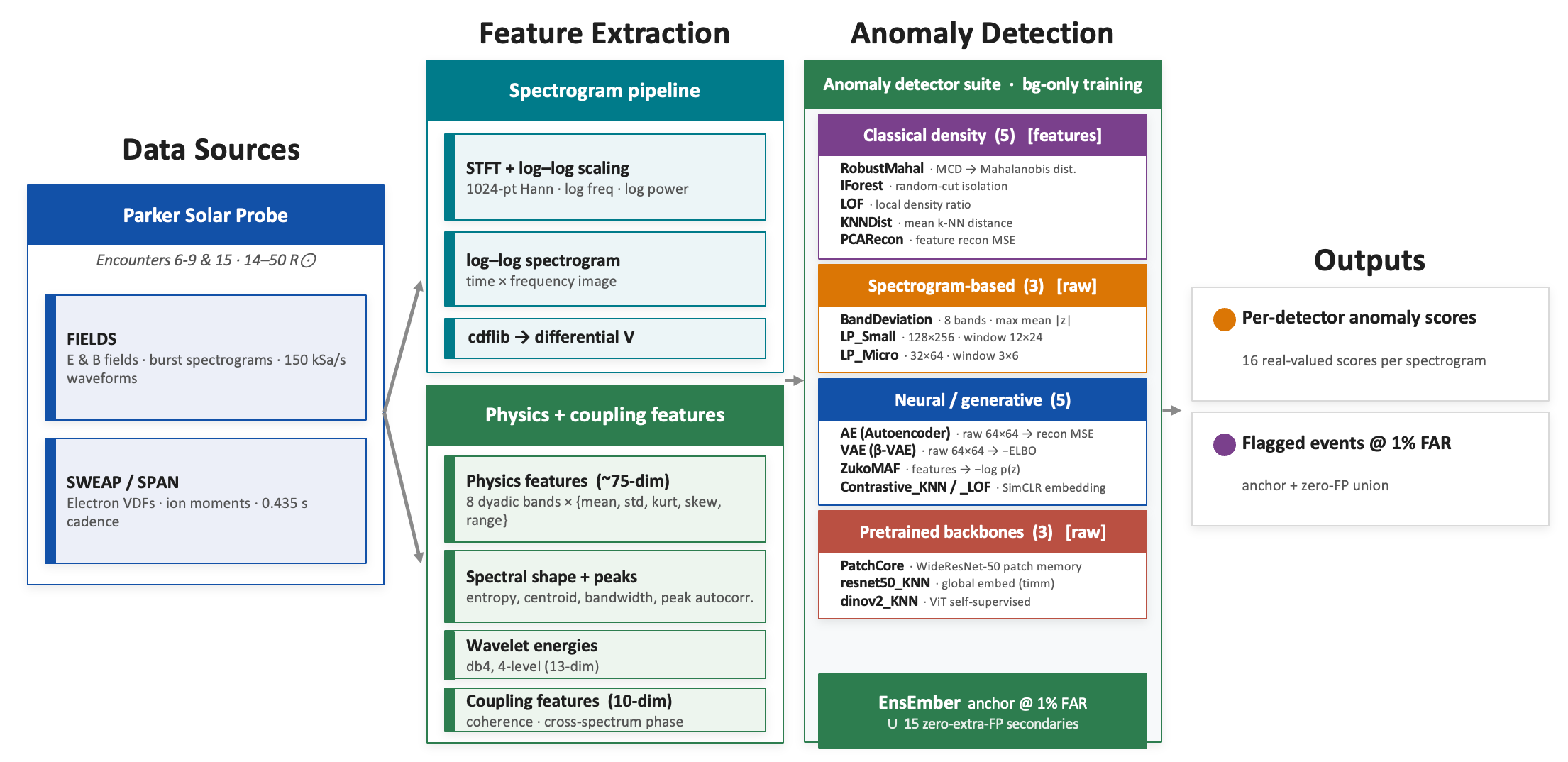}
\caption{\textsc{Ember} pipeline overview. PSP FIELDS DBM bursts are converted to log–log Fourier spectrograms and, in parallel, a physics+coupling feature bank is extracted. Sixteen background-only anomaly detectors, grouped into four families, are scored per spectrogram. The EnsEmber anchor is a differential-evolution-optimized weighted ensemble calibrated at a 1\% false-alarm rate; the zero-extra-FP union adds the 15 secondary detectors at thresholds that contribute no additional false positives on the calibration set. SWEAP/SPAN electron and ion moments provide independent heating diagnostics (Section 5) and are not used in the detection pipeline.}
\label{fig:workflow}
\end{figure}

\textsc{Ember} is organized as a sequence of data-preparation and machine-learning stages implemented in Python and released as an open-source package (\url{https://github.com/asasli/EMBER}). A schematic of the end-to-end workflow is shown in Figure~\ref{fig:workflow}. Raw DBM voltage bursts are converted to log-scaled Fourier spectrograms (Section~\ref{sec:pre}), a suite of 16 complementary detectors is trained on background-only spectrograms (Section~\ref{sec:detectors}), and individual detector
scores are combined through a calibrated ensemble (Section~\ref{sec:ensemble}).

\subsection{Preprocessing and spectrogram representation}
\label{sec:pre}

For each burst window, \textsc{Ember} (i) reads the CDF file with \texttt{cdflib}; (ii) extracts the voltage time series (DVAC directly, or computed differentially from VAC); (iii) converts TT2000 timestamps to UTC; (iv) computes a short-time Fourier transform using a 1024-point Hanning window; and (v) applies logarithmic scaling to both the frequency and power axes. The dual log-scaling is not cosmetic: modulated IAW signatures manifest as a low-frequency (${\sim}1$--$20\,\mathrm{Hz}$) envelope modulating a higher-frequency (${\sim}100$--$1000\,\mathrm{Hz}$) carrier, and these two bands span roughly three decades in frequency and more than six decades in power. Linear axes compress the modulating envelope and the high-frequency fine structure into unresolved bands, whereas log--log axes render the characteristic ladder-of-sidebands morphology of triggered and modulated IAWs clearly visible. The log--log spectrogram is the canonical input representation for all downstream stages.

\subsection{Physics-motivated feature bank}
\label{sec:features}

In addition to the raw spectrogram, \textsc{Ember} computes a bank of physics-motivated features grouped into two families. \emph{Physics features} summarize the frequency-resolved power envelope: band-integrated power in dyadic frequency bands spanning the expected ion-acoustic range, spectral flatness, spectral centroid and bandwidth, roll-off frequency, and estimates of the local plasma and ion plasma frequencies scaled by the nominal density and magnetic field during each burst. \emph{Coupling features} target the bi-scale modulation signature: cross-band power ratios (low-frequency to high-frequency band), amplitude-modulation depth and coherence estimated from the analytic envelope of the high-frequency band, and column-to-column spectrogram similarity metrics that detect persistent narrowband carriers. A \texttt{PhysicsAugmenter} generates label-preserving augmentations of these features (frequency jitter, amplitude rescaling, additive colored noise at levels consistent with the measured instrument noise floor) during training of the detectors described in Section~\ref{sec:detectors}. The feature bank is curated by \texttt{summarize\_feature\_discrimination}, which ranks
features by background-versus-anomaly separability on a training fold and retains only those whose rank-based separation exceeds a preregistered threshold.

\subsection{Detector suite}
\label{sec:detectors}

\textsc{Ember} implements 16 anomaly detectors that operate either directly on the log--log spectrograms or on the physics/coupling feature bank of
Section~\ref{sec:features}. All detectors are trained exclusively on background spectrograms ($N_{\rm train}=396$) and calibrated on a disjoint held-out set of 100 background spectrograms ($N_{\rm cal}=100$; Section~\ref{sec:ensemble}). No anomaly labels are used during training of any individual detector. We group the detectors into four families.

\subsubsection{Physics-motivated detectors.}
The \texttt{BandDeviation} models, for each frequency band in the physics feature bank, the background distribution of integrated band power as a heavy-tailed
density and assigns an anomaly score equal to the maximum over bands of the deviation (in background-standard-deviation units) of the observed band power. It is explicitly targeted at narrowband electrostatic intensification below the local plasma frequency, the canonical TIAW signature \citep{mozer2,vech}. The \texttt{LocalPatch} tiles each spectrogram into overlapping time--frequency patches, builds a background memory bank of patch-level feature vectors, and scores a test spectrogram by the distance of its worst patch to the bank; it is designed to catch localized coupled-wave structures that do not perturb the globally integrated power but alter the joint time--frequency distribution. Two configurations are evaluated (\textit{small} and \textit{micro} patch grids) to probe different spatial scales of the wave modulation pattern.

\subsubsection{Classical density-based detectors on the feature bank.}
Five classical detectors operate on the physics/coupling feature bank. \textbf{Isolation forest} \citep{Liu2008} partitions the feature space with random axis-aligned cuts and assigns anomaly scores proportional to the path length required to isolate a sample. \textbf{Local outlier factor} \citep{Breunig2000} compares the local reachability
density of each sample with that of its $k$ nearest neighbors. \textbf{$k$-NN distance} scores each sample by its distance to its $k$-th nearest neighbor in the background training set \citep{Ramaswamy2000}. \textbf{Robust Mahalanobis distance} fits a minimum-covariance-determinant estimate \citep{Rousseeuw1999} of the background mean $\boldsymbol{\mu}$ and covariance $\boldsymbol{\Sigma}$ and scores a test vector $\mathbf{x}$ by
\begin{equation}
s_{\rm Mah}(\mathbf{x})
    \;=\; \sqrt{(\mathbf{x}-\boldsymbol{\mu})^{\top}
                \boldsymbol{\Sigma}^{-1}
                (\mathbf{x}-\boldsymbol{\mu})}.
\label{eq:mahalanobis}
\end{equation}
\textbf{PCA reconstruction error} projects each sample onto the leading principal components of the background feature matrix and scores it by the norm of the residual in the discarded subspace; large residuals indicate features absent from the background manifold.

\subsubsection{Deep detectors on the spectrograms.}
Four deep detectors operate directly on the log--log spectrograms. The \textbf{convolutional autoencoder} (AE) is trained to reconstruct background
spectrograms by minimising the mean squared error; anomalies are scored by the 95th-percentile per-pixel reconstruction error to avoid dilution by uninformative
regions. The \textbf{$\boldsymbol\beta$-variational autoencoder} (\texttt{SpectrogramVAE};
\citealt{Higgins2017}) learns a disentangled latent representation of background spectrograms by maximising the $\beta$-weighted evidence lower bound (ELBO)
\begin{equation}
\mathcal{L}_{\beta\text{-VAE}}(\phi,\theta;\mathbf{x})
\;=\;
\mathbb{E}_{q_\phi(\mathbf{z}\mid\mathbf{x})}
    \!\left[\log p_\theta(\mathbf{x}\mid\mathbf{z})\right]
\;-\;
\beta\,D_{\mathrm{KL}}\!\left(q_\phi(\mathbf{z}\mid\mathbf{x})
                               \,\big\|\,p(\mathbf{z})\right),
\label{eq:bvae}
\end{equation}
and assigns an anomaly score equal to the negative ELBO at test time.  A \textbf{masked autoregressive flow} (MAF) \citep{Papamakarios2017} 
implemented with the \textsc{Zuko} library \citep{Zuko2022}  (\texttt{ZukoMAF}) learns a bijective map from the background physics-feature distribution to a standard Gaussian and scores test samples by $-\log p(\mathbf{z})$ where $\mathbf{z} = f_\psi(\mathbf{x})$, 
providing a calibrated density-based score complementary to the reconstruction-based AE and ELBO-based $\beta$-VAE.
A \textbf{contrastive encoder} \citep{Chen2020} is pre-trained self-supervised on augmented pairs of background spectrograms; the resulting low-dimensional embedding is scored with both $k$-NN distance (\texttt{Contrastive\_KNN}) and local outlier factor 
(\texttt{Contrastive\_LOF}) to give two complementary anomaly detectors.

\subsubsection{Pretrained-backbone detectors.}
Three detectors leverage large pretrained vision models without any fine-tuning on
PSP data. \texttt{PatchCore} \citep{Roth2022} extracts mid-level patch features from a frozen ImageNet-pretrained WideResNet-50 \citep{Zagoruyko2016}, builds a coreset memory bank of background patch features via greedy subsampling, and scores a test spectrogram by the distance of its worst patch to the bank. A \texttt{ResNet-50} \citep{He2016} global embedding followed by $k$-NN scoring
exploits deeper semantic features of the same ResNet family at full-image resolution. A \texttt{DINOv2} \citep{Oquab2023} vision transformer global embedding followed by $k$-NN scoring provides a self-supervised, attention-based representation that captures long-range frequency correlations invisible to convolutional detectors. Although none of these backbones has seen solar-wind data, their low- and mid-level features transfer well to spectrograms and consistently outperform the deep detectors trained from scratch (Table~\ref{tab:detectors}).

\subsection{Ensembling and FAR-calibrated operating point}
\label{sec:ensemble}

Every detector in Section~\ref{sec:detectors} produces a real-valued anomaly score. To combine detectors with incompatible score scales we apply rank normalisation to each detector's scores, mapping them to $[0,1]$ based on their empirical rank within the held-out calibration set of $N_{\rm cal}=100$ background spectrograms. Rank normalisation is computed once per detector and never refit on anomaly labels.

We denote by \texttt{EnsEmber} the weighted rank-normalised ensemble whose weights are optimised by differential evolution \citep{Storn1997} as implemented in
\textsc{SciPy} \citep{Virtanen2020} to maximise the true-positive rate (TPR) at a fixed false-alarm rate (FAR) of $1\,\%$, evaluated on the held-out calibration set. The ensemble pool consists of the six detectors that individually achieve the highest AUC on the calibration set: \texttt{PatchCore}, \texttt{resnet50-KNN}, \texttt{ZukoMAF}, \texttt{DINOv2-KNN}, \texttt{PCARecon}, and \texttt{KNNDist}. This operating point corresponds to at most one false positive per 100 held-out background spectrograms ($\text{FAR}=1/100=1\,\%$).

Beyond the ensemble, we construct a \emph{zero-extra-FP union} that augments \textsc{EnsEmber} with every secondary detector whose detection threshold can be set strictly above its maximum calibration-set score, guaranteeing no additional false positives from the calibration set. By the monotonicity of anomaly detection -- lowering a threshold can only add detections, never remove them -- the zero-extra-FP threshold is the loosest possible threshold that preserves the FAR budget. All 15 remaining detectors contribute additional anomaly detections under this rule, raising the union TPR from $61.9\,\%$ (26/42 events for \textsc{EnsEmber} alone) to $92.9\,\%$ (39/42 events) while the FAR remains fixed at $1.00\,\%$. We note that FAR value is approximate due to limited background statistics. Label-1 recovery improves from $36.8\,\%$ to $84.2\,\%$ (16/19 events) and Label-2 recovery reaches $100\,\%$ (23/23 events). The three events that remain undetected (\#10, \#258, \#410) score below the maximum calibration-set score for every detector and therefore cannot be recovered without increasing the FAR budget. Importantly, the anomaly labels are used only to tune the \textsc{EnsEmber} weights -- a standard practice when anomaly examples are available but scarce -- and never to train any individual detector.

\subsection{Coincident electron-heating diagnostics}
\label{sec:heating-method}

For every \textsc{Ember}-flagged interval we extract coincident SWEAP/SPAN electron velocity distribution functions (VDFs) and compute core perpendicular and parallel
temperatures, $T_\perp$ and $T_\parallel$. The analysis is restricted to the core population and excludes low-energy bins ($<34\,\mathrm{eV}$) to avoid spacecraft contamination. VDFs are rotated into the magnetic-field-aligned frame using FIELDS magnetometer data and shifted to the plasma rest frame using bulk solar-wind velocities from SPAN-Ai. A nonlinear least-squares fit to a bi-Maxwellian distribution
\citep{halekas,berc}
\begin{equation}
f_c(v_\perp,v_\parallel)
\;=\;
n_c \left(\frac{m_e}{2\pi k_{\rm B}}\right)^{\!3/2}
\frac{1}{T_\perp\sqrt{T_\parallel}}\,
\exp\!\left[
  -\frac{m_e v_\perp^2}{2k_{\rm B}T_\perp}
  -\frac{m_e(v_\parallel - v_d)^2}{2k_{\rm B}T_\parallel}
\right]
\label{eq:bimax}
\end{equation}
yields the core electron temperatures, where $v_d$ is the field-aligned drift velocity. Core temperatures are compared against the adiabatic cooling expectation $T_\perp \propto R^{-4/3}$ \citep{mozercore} to identify intervals of excess heating, and core electron-to-proton temperature ratios $T_e/T_i$ are computed using SPAN-Aiion moments over the same intervals. Critically, these electron-heating products are \emph{not} used anywhere in the detection pipeline of Sections~\ref{sec:detectors}--\ref{sec:ensemble}; they provide an independent physical test of the significance of \textsc{Ember}'s flags.

\section{Detection performance}
\label{sec:results}

At the $1\,\%$ FAR operating point, the \textsc{EnsEmber} ensemble, augmented by its zero-extra-FP union with the 15 strongest secondary detectors, recovers \textbf{93\,\%} of the 42 anomalous events (i.e.\ 39 of 42 events flagged) in the
hand-labeled catalog of Section~\ref{sec:catalog}, at a cost of one false positive in the 100-sample held-out background set. Table~\ref{tab:detectors} summarizes the single-detector and ensemble performance at the same operating point.

\begin{table}[ht]
\centering
\caption{Per-detector performance on the held-out calibration set
         ($N_{\mathrm{cal}}=100$ background samples, 42 anomalies: L1\,$=\,19$, L2\,$=\,23$). The anchor threshold is set to give at most one false positive on $N_{\mathrm{cal}}$ (FAR\,${=}\,1\,\%$); secondary detectors are thresholded at zero false positives on $N_{\mathrm{cal}}$. The union row combines the \textsc{EnsEmber} anchor with the 15 zero-FP secondaries at no extra false-alarm cost. Detector names match the \textsc{Ember} implementation. L1 TPR and L2 TPR are the recovery rates on the two physical label classes 
(L1: modulated / TIAW-like, $n=19$; L2: FDIAWs with weak heating, $n=23$).}
\label{tab:detectors}
\begin{tabular}{@{}lcccc@{}}
\toprule
Detector & AUC & TPR\,@\,1\,\% & L1 TPR & L2 TPR \\
\midrule
\multicolumn{5}{@{}l}{\textit{Spectrogram-based}} \\
\quad BandDeviation                          & 0.775 &  0 &  0 &   0 \\
\quad LP\_Small  (LocalPatch, 128$\times$256) & 0.886 &  2 &  5 &   0 \\
\quad LP\_Micro  (LocalPatch, 32$\times$64)   & 0.880 &  2 &  5 &   0 \\
\midrule
\multicolumn{5}{@{}l}{\textit{Classical density-based (feature bank)}} \\
\quad IForest                                 & 0.931 & 43 & 11 &  70 \\
\quad LOF                                     & 0.783 &  2 &  0 &   4 \\
\quad KNNDist                                 & 0.936 & 45 & 16 &  70 \\
\quad RobustMahal                             & 0.922 & 24 &  5 &  39 \\
\quad PCARecon                                & 0.948 & 45 & 16 &  70 \\
\midrule
\multicolumn{5}{@{}l}{\textit{Neural / generative}} \\
\quad AE (convolutional autoencoder)          & 0.935 & 36 & 16 &  52 \\
\quad VAE ($\beta$-VAE)                        & 0.886 &  0 &  0 &   0 \\
\quad ZukoMAF (masked autoregressive flow)    & 0.971 & 48 & 11 &  78 \\
\quad Contrastive\_KNN                         & 0.858 & 14 & 11 &  17 \\
\quad Contrastive\_LOF                         & 0.780 & 12 & 11 &  13 \\
\midrule
\multicolumn{5}{@{}l}{\textit{Pretrained backbones}} \\
\quad \textsc{PatchCore} (WideResNet-50)       & 0.986 & 50 & 16 &  78 \\
\quad resnet50\_KNN                            & 0.982 & 43 & 21 &  61 \\
\quad dinov2\_KNN                              & 0.970 & 55 & 58 &  52 \\
\midrule
\multicolumn{5}{@{}l}{\textit{Ensemble}} \\
\quad EnsEmber\_equal (equal weights)          & 0.958 & 48 & 21 &  70 \\
\textbf{EnsEmber} (rank-norm., DE-optimised)   & \textbf{0.971} & \textbf{62} & \textbf{37} & \textbf{83} \\
\midrule
\textbf{EnsEmber} $\cup$ \textbf{15 zero-FP secondaries}
                                               & ---   & \textbf{93} & \textbf{84} & \textbf{100} \\
\bottomrule
\end{tabular}
\end{table}

\begin{figure}[t]
\centering
\includegraphics[scale=0.32]{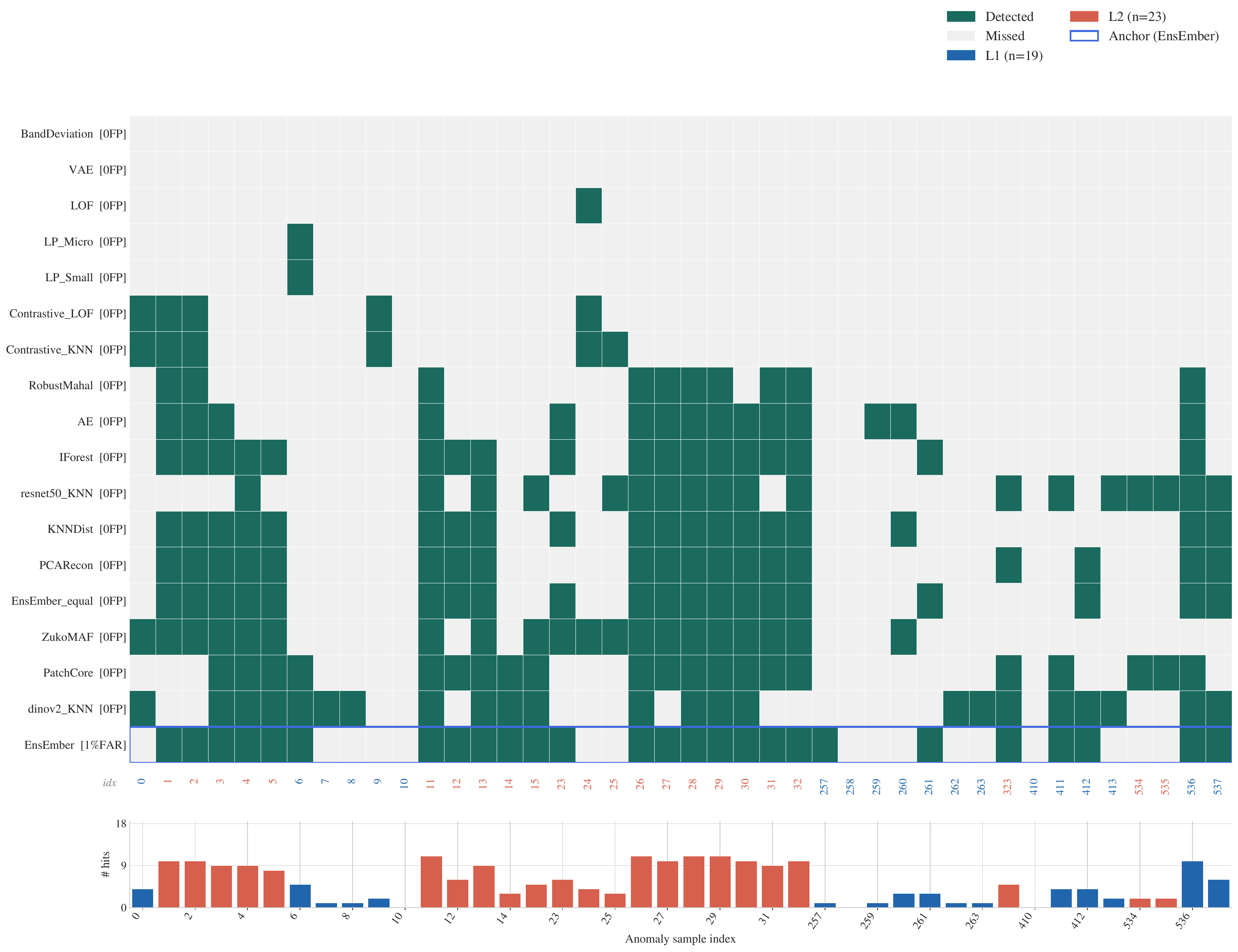}
\caption{Detection map: per-event recovery of the \textsc{Ember} detector suite
         at $1\,\%$ FAR.
         \textit{Columns}: 42 anomalous events (Label~1 in blue, Label~2 in red).
         \textit{Rows}: detectors labelled with their threshold rule
         (\texttt{[1\%FAR]} for the \textsc{EnsEmber} anchor;
          \texttt{[0FP]} for all secondaries).
         Filled cells: detector flags event.
         \textit{Bottom}: number of detectors flagging each event.}
\label{fig:detection-map}
\end{figure}

Figure~\ref{fig:detection-map} visualizes \emph{which} events are recovered. Each column is one of the 42 anomalous events (Label~1 in blue, Label~2 in red); each row is a detector, ordered by sensitivity. Filled cells indicate that the corresponding detector flagged that event at the $1\,\%$ FAR operating point. The bar chart at the bottom counts the number of detectors flagging each event. Two trends are visible. First, Label~2 events (FDIAWs with weak heating) are flagged by more detectors on average than Label~1 events (modulated / coupled TIAW-like waves, associated with strong core-electron heating), consistent with their higher morphological complexity and the $100\,\%$ L2 recovery of the union detector. Second, the top-performing deep detectors (\textsc{PatchCore}, MAF, contrastive encoder) and the physics-motivated detectors cover largely complementary subsets of events, which is why their rank-normalized union under a fixed FAR budget outperforms any individual method. Three events (\#10, \#258, \#410) remain undetected by all detectors at zero extra false positives; they represent the irreducible hard cases at this operating point.

\begin{table}[ht]
\centering
\caption{Calibration-set union analysis results.
         The anchor (\textsc{EnsEmber}) is run at $1\,\%$ FAR on the held-out
         calibration set ($N_\text{cal}=100$).
         Secondary detectors are added at a zero false-positive threshold;
         their union contributes additional true positives at no extra FAR cost.}
\label{tab:cal_union}
\begin{tabular}{@{}lcc@{}}
\toprule
\textbf{Configuration} & \textbf{Metric} & \textbf{Value} \\
\midrule
\multicolumn{3}{@{}l}{\textit{Setup}} \\
\quad Calibration set size  & $N_\text{cal}$  & 100 samples \\
\quad Anomalies             & $N_\text{anom}$ & 42\quad(L1:\,19,\ L2:\,23) \\
\midrule
\multicolumn{3}{@{}l}{\textit{Anchor: \textsc{EnsEmber} @ $1\,\%$ FAR}} \\
\quad True positives        & TP  & 26\,/\,42\quad(62\,\%) \\
\quad False positives       & FP  & \ \ 1\,/\,100\quad(1\,\%) \\
\midrule
\multicolumn{3}{@{}l}{\textit{Union: anchor $+$ 15 zero-FP secondaries}} \\
\quad Free TP gain          & $\Delta$TP                     & $+13$ at 0 extra FP \\
\quad True positives        & TP                             & 39\,/\,42\quad(93\,\%) \\
\quad \ \ L1 recovery       & TPR$_{\text{L1}}$              & 16\,/\,19\quad(84\,\%) \\
\quad \ \ L2 recovery       & TPR$_{\text{L2}}$              & 23\,/\,23\,(100\,\%) \\
\quad False positives       & FP                             & \ \ 1\,/\,100\quad(1.00\,\%) \\
\quad FAR                   &                                & 1.00\,\% (unchanged) \\
\midrule
\quad Missed events         &                                & 3\quad(\#10, \#258, \#410) \\
\bottomrule
\end{tabular}
\end{table}

\section{Coincident core-electron heating}
\label{sec:heating}

\begin{figure}[t]
\centering
\includegraphics[scale=0.29]{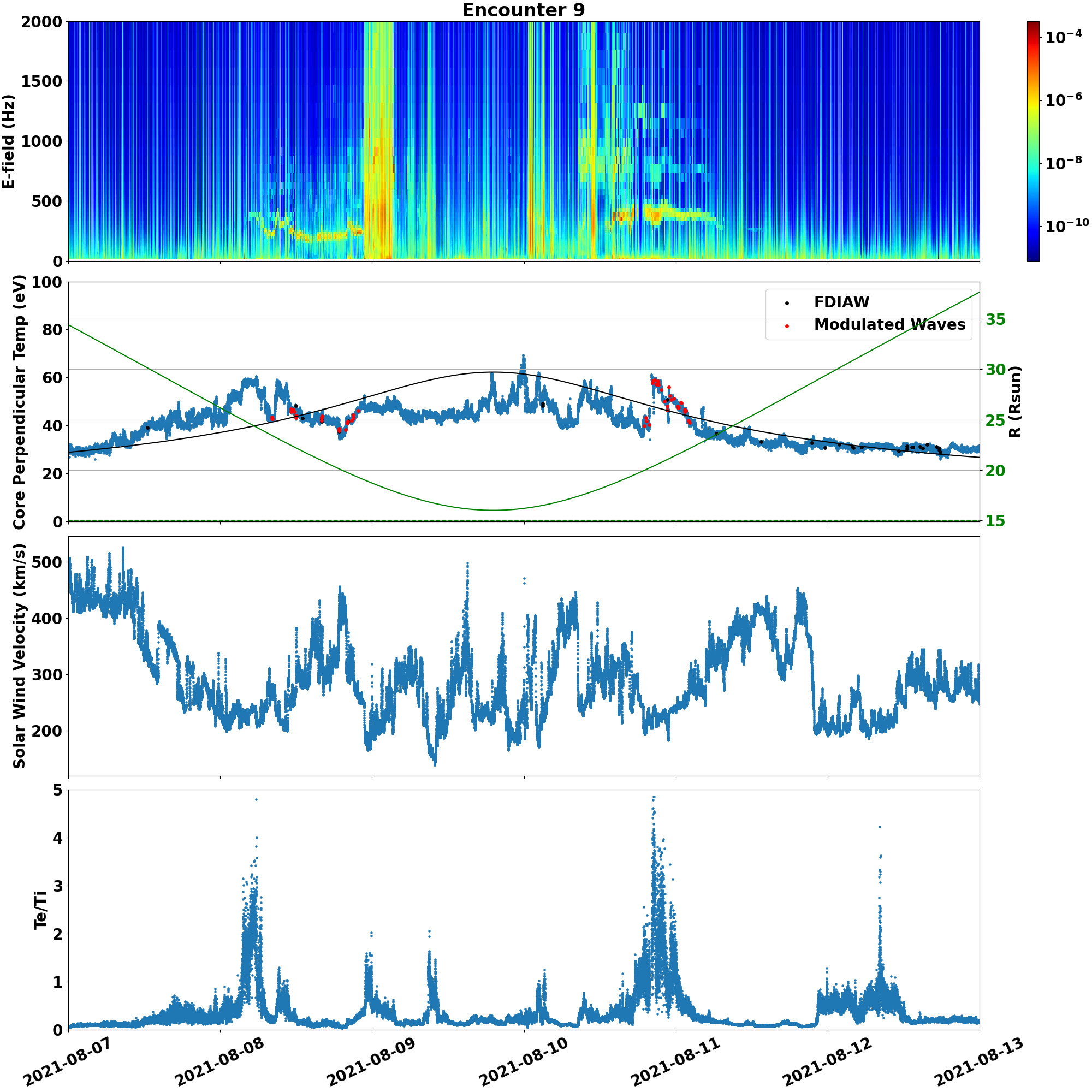}
\caption{PSP Encounter~9 overview: wave activity and coincident core-electron heating. Elevated core electron temperatures can be seen in the presence of modulated waves despite relatively stable solar wind velocities.}
\label{fig:encounter9}
\end{figure}

Having established that \textsc{Ember} flags 93\% of the catalog events at 1\% FAR, we now show that the flagged intervals reproduce the electron-heating phenomenology of earlier manual studies. Figure~\ref{fig:encounter9} summarizes PSP Encounter~9. The top panel shows the FIELDS DFB AC electric-field spectrogram; the middle panel shows the core perpendicular electron temperature $\Tperp$ (blue), with the adiabatic cooling expectation $\Tperp \propto R^{-4/3}$ overplotted as a black line; the bottom panel shows the ratio $\Te/\Ti$. The third panel shows the solar wind velocity as measured by PSP in-situ. The core electron temperatures have a negative correlation with solar wind velocity as the faster solar wind consists of cooler electrons. Black markers indicate frequency-dispersed IAWs, some of which are included in the hand-labeled catalog (Label~2); colored markers indicate modulated-wave events (Label~1). Intervals containing Label~1 events exhibit $\Tperp$ well above the adiabatic expectation and $\Te/\Ti \gtrsim 2$, indicating preferential electron heating. Label~2 intervals show markedly weaker deviations.

\begin{figure}[t]
\centering
\includegraphics[scale=0.2]{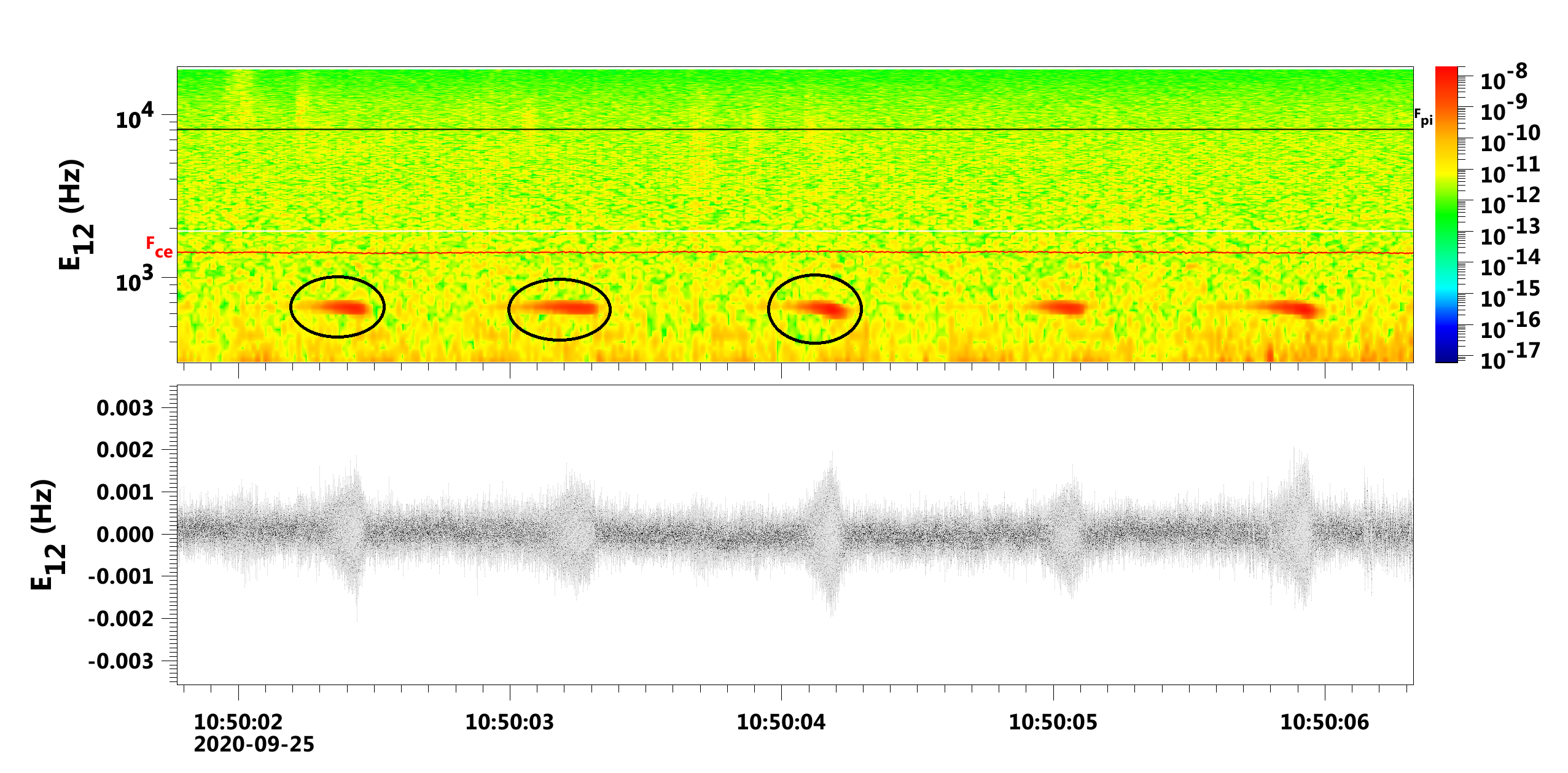}
\caption{Example of modulated waves seen during Encounter 6. These events coincided with elevated core electron temperatures.}
\label{fig:encounter6}
\end{figure}

Restricting attention to the set of intervals flagged by \textsc{Ember}, we recover this same qualitative picture: the flagged Label~1 events lie systematically above the adiabatic profile in $\Tperp$ and at elevated $\Te/\Ti$, while the flagged Label~2 events show weaker excursions. Because $\Tperp$, $\Tpar$, and $\Te/\Ti$ are \emph{not} inputs to any stage of the \textsc{Ember} detection pipeline, this recovery is an independent physical test that the log--log spectrogram anomalies flagged by \textsc{Ember} carry the kinetic heating signature targeted by the study. Comparison between Encounter~6 ($\sim 21\text{--}40\,\Rsun$) and Encounter~15 ($\sim 12\text{--}35\,\Rsun$) shows that modulated-wave occurrence per burst hour and the associated $\Tperp$ excess both are significantly enhanced at smaller heliocentric distances (Encounter 15) compared to larger distances (Encounter 6), consistent with the picture in which modulated IAWs are an intrinsic feature of the near-Sun solar wind and contribute preferentially to core-electron heating in the inner heliosphere.
\section{Discussion}
\label{sec:discussion}

Supervised classification on spectrograms --- for example, a convolutional neural network trained end-to-end on the catalog --- is the obvious alternative to the approach taken here. We deliberately avoid it for three reasons. First, with only 42 anomalous events across two morphological classes, any high-capacity supervised model faces severe overfitting risk and would likely fail to generalize beyond the catalog. Second, a large fraction of the scientific value of a catalog of wave-driven heating events lies in the long tail of unusual events that a supervised classifier would be trained to deprioritize. Third, background only training provides a natural guarantee against label-contamination artifacts: detectors never see anomalies during training, so reported TPR@FAR numbers are not inflated by memorization.

The multi-detector structure addresses the morphological diversity of the targets. Band-integrated power, local time--frequency patch structure, global feature-bank outlierness, reconstruction quality, and density under a learned generative model each probe different aspects of the spectrogram. No single detector dominates across both label classes (Figure~\ref{fig:detection-map}): the most unusual events tend to be caught by multiple detectors, but the marginal events caught by only one or two are precisely the cases where the diversity of the suite pays off. Rank-normalized ensembling with FAR-constrained weight optimization is, for this reason, the natural aggregation strategy.

The three labeled events missed by the union at 1\% FAR (\#10, \#258, \#410, all L1) cluster in two regimes. The first is low-amplitude modulated activity with weakly coherent envelopes; these events sit near the decision boundary and would be recovered at higher recall settings at the cost of additional false positives. The second is events embedded in intervals of elevated broadband background turbulence, where the anomaly score is suppressed because the background itself is already unusual.

A major motivation for releasing \textsc{Ember} as a modular pipeline is that its feature layer can be replaced or augmented without changing the detector suite or the ensemble. The \textsc{Surya} Helio Foundation Model, developed by NASA's Heliophysics Division and the IMPACT Office, ingests multichannel full-disk observations from the Solar Dynamics Observatory (SDO) \citep{pesnell}, including EUV imagery from the Atmospheric Imaging Assembly and magnetograms from the Helioseismic and Magnetic Imager, and learns high-dimensional latent representations of the global coronal magnetic state through self-supervised pretraining. These embeddings provide a compact, data-driven summary of the Sun's magnetic configuration without requiring explicit physical labels during training.

In a follow-up phase of this work we will concatenate \textsc{Surya} embeddings (for the SDO frame contemporaneous with each PSP burst) to the \textsc{Ember} physics/coupling
feature bank. This enables two scientific extensions that are not possible with spectrogram features alone: (i) supervised prediction of PSP-measured heating outcomes ($\Te/\Ti$ enhancement, $\Tperp$ deviation from adiabatic cooling, modulated-wave occurrence probability) from purely remote-sensing coronal inputs, and (ii) unsupervised clustering of the joint \textsc{Surya}--PSP latent space to discover distinct coronal heating regimes, which can be mapped back onto full-disk SDO imagery. The detection performance reported here establishes the in-situ side of that pipeline as mission-validated and ready for integration.

\section{Conclusions}
\label{sec:conclusions}

We have presented \textsc{Ember}, an open-source machine-learning pipeline that processes Parker Solar Probe FIELDS DBM burst-mode data into log--log Fourier spectrograms and applies a multi-detector, background-only anomaly-detection suite.
The suite combines physics-motivated detectors (\texttt{BandDeviation} and \texttt{LocalPatch}), classical outlier detectors on the physics/coupling feature bank (RobustMahal, IForest, LOF, k-NN distance, and PCA reconstruction), and deep detectors (autoencoder, $\beta$-VAE, normalizing flow, contrastive encoder, and \texttt{PatchCore} with a frozen \texttt{WideResNet}-50 backbone). Scores are rank-normalized and combined through an ensemble whose weights are optimized by differential evolution to maximize TPR at a fixed 1\% false-alarm rate. On a catalog of 496 background and 42 anomalous DBM spectrograms from PSP Encounters~6 through~9 and~15, the \textsc{Ember} ensemble recovers 93\% of the anomalous events at 1\% FAR. Coincident SWEAP/SPAN electron diagnostics
--- which are not used anywhere in the detection pipeline --- confirm that flagged intervals exhibit core $\Tperp$ above adiabatic cooling expectations and elevated $\Te/\Ti$, reproducing the preferential-heating phenomenology of earlier manual studies. The pipeline is modular and reproducible, its feature layer is designed to be straightforwardly augmented with foundation-model embeddings, opening a path toward connecting global coronal magnetic context to in-situ kinetic heating signatures in the near-Sun solar wind.

%
%

\section*{Open Research Section}
All PSP FIELDS and SWEAP Level-2 data used in this study are publicly available from NASA CDAWeb (\url{https://cdaweb.gsfc.nasa.gov}) and Berkeley SSL
(\url{https://research.ssl.berkeley.edu/data/psp/}). The \textsc{Ember} pipeline, including the multi-detector suite, the ensemble optimizer, and the reporting/plotting helpers used
to produce Figures~\ref{fig:workflow}--\ref{fig:encounter6} and Table~\ref{tab:detectors}, is publicly available at \url{https://github.com/asasli/EMBER} under a permissive
open-source license. The labeled catalog of 496 background and 42 anomalous DBM spectrograms and the trained detector and ensemble weights used in this paper will be archived on Zenodo with a DOI upon acceptance.

\acknowledgments

A.S. and M.W.C. acknowledge support from the National Science Foundation with grant numbers PHY-2117997, PHY-2308862 and PHY-2409481.

\bibliography{agusample}

\end{document}